# SUSTAINABLE DEVELOPMENT THROUGH A MOBILE APPLICATION FOR A COMMUNITY CLINIC

Martina A. Clarke, University of Nebraska-Omaha, USA, martinaclarke@unomaha.edu

Sajda Qureshi, University of Nebraska-Omaha, USA, squreshi@unomaha.edu

Timi Barone, University of Nebraska-Omaha, USA, tbarone@unomaha.edu

John R. Windle, University of Nebraska, USA, jrwindle@unmc.edu

**Abstract:** Implementing Information and Communication Technology (ICT) solutions can alleviate pressing problems in society and are a central component of sustainable development. Often, healthcare addresses the symptoms without approaching the socioeconomic limiters that can lead to reduced individual economic freedoms from receiving healthcare. This paper investigates the question: How can technology and training interventions enable clinicians to offer care that addresses the socioeconomic limitations of their patients? This paper observes the implementation of a mobile app designed to offer people who cannot access health resources in Omaha, Nebraska, a city in the Midwestern United States. This study follows the design science and action research approach, with clinicians participating in developing the mobile app. As a result of COVID, patients no longer have access to the free clinic because it was shut down. The app is available to the broader community needing basic resources to stay healthy. Through sets of application revisions and observations of usage, this paper arrives at insights into how such applications can support multi-ethnic and underserved communities. The contribution of this paper is to provide contextually specific and rich descriptions of how to implement sustainable ICT solutions to meet the information needs of patients in underserved communities.

## 1. INTRODUCTION

The concept of development can be understood from three different scopes 1) generic development, which is progressive change as a society, 2) geography-specific development, which is progressive change in a developing country, and 3) agenda-specific development, which is continuous changes in a developing country where societies are divided into government, community, and the private sector (Heeks, 2017). Information and Communications Technologies for Development (ICT4D) is about social and technical research on the effects of Information Technology (IT) on economic, social, and human development. The eradication of poverty is the primary pillar of sustainable development and there is conclusive evidence that health directly correlates with the rates of poverty in an observed community. Attending to these aspects together are the precursor to development in a community. In Omaha, Nebraska, a city in the Midwestern United States, it is estimated that 68,334 people or 15.1 percent of the entire population live below the poverty line (Office of Assistant Secretary for Planning and Evaluation, 2020); this means that poverty rates in Omaha are 2.8 percent higher than the national average of 12.3 percent. Based on the Nebraska Health Disparities Report, American Indians report the highest number of socio-economic indicators of disparity, followed by African Americans (Anthone, Medinger, Rodriguez 2020).

The goal of poverty eradication is to reinvigorate the human element of society and to further the creation of opportunity. The most direct method for enticing human and social development is for community members to participate in nonprofits that support the health and wellbeing of the struggling and underprivileged, where economic development is mostly non-existent. The pandemic has increased the need for community resources that can help people stay healthy. These factors heighten the necessity to provide accessible healthcare, especially in a non-profit form.

Sustainable Development is "meeting the needs of the people without affecting the future." Sustainable Development falls under the chronology of development paradigms. Development paradigms started from Modernization in the 1950s, where development, growth, technology, and the transfer of ideas took place. In the 1960s, development focused on dependency which are core-periphery positions within the world system and countries breaking away from an exploitative world system. In the 1970s basic needs of an individual focused on food, clothes, shelter. In the 1980s, Neoliberalism began to take hold, which focused mainly on markets. In the 1990s the human development concept started which focuses on gender, education, poverty and where humans have freedom to live and do what they want to do. In the 2000s, the sustainable development phase started, which focuses on "meeting current needs without compromising future needs, especially the environment (Heeks, 2017)."

Socio-economic status is seen to be linked to health. Adler et al. (1994) contend that individuals living in better socio-economic conditions enjoy better health than those living in lower socio-economic conditions. They reported on a model that illustrates a directly inverse correlation between morbidity rates and socio-economic status. They found that the lower the socio-economic status of individuals the higher the percentage of people were diagnosed with osteoarthritis, chronic diseases, hypertension, and cervical cancer. Additionally, education, income, and life expectancies are key factors that lead to different socio-economic conditions (Adler et al., 1994). Roztocki and Weistroffer (2016) define socio-economic development to be "a process of change or improvements in social and economic conditions as they relate to an individual, an organization, or a whole country (p.542)."

In this paper the concept of socio-economic development is seen as a combination of social factors such as healthcare, wellbeing, and economic factors, such as, income, and assets. To be healthy and lead the lives they choose to live, people need access to social and economic resources (Siahpush, 2019; Ramos, 2020; Clarke, 2020). This paper aims to investigate the question: How can technology and training interventions enable clinicians to offer care while addressing the socio-economic limitations of their patients? This paper observes the implementation and use of a mobile app for a student led free clinic that serves patients who are uninsured and unable to pay for their healthcare. It follows an "Information Technology Therapy" process in which clinicians are supported with the mobile app and training to help them better manage the care of their patients. Through sets of app revisions and observations of usage, this paper arrives at insights into how such applications can support multi-ethnic and underserved communities. The contribution of this paper is to provide contextually specific and rich descriptions of how to implement sustainable Information and Communications Technologies (ICT) solutions to meet the information needs of patients in underserved communities.

## 1.1. Sustainable Development and Health Inequities

ICT support health and wellbeing by standardizing and disseminating the information essential to the maintenance of a community's people and their health, creating efficiencies, and improving the interoperability of collected data. By increasing accessibility to healthcare and its degree of quality through ICTs, the time spent between inhibiting health issues and being a more productive member of the community is minimized (Jones, 2018). Qureshi & Xiong (2021) found a relationship between mobile phone and internet usage on the health and wellbeing of a population. They found that inequalities in education and income moderate the positive relationship between the mobility effect for health and human development. In some cases, higher health standards can be seen to be attributed to better education. This means that people with higher education can access the resources, including information, they need to stay healthy. They are also seen to have higher economic outcomes within a local community and strengthens the community from within.

Equitable healthcare provision entails socio-economic development (Roztocki and Weistroffer 2016, Qureshi & Xiong 2021). Accessibility to healthcare and ICT-based solutions that are involved with, and help facilitate healthcare accessibility are needed in these communities. ICTs are efficient and cost-effective solutions to the management of health needs and for assisting the duties of healthcare professionals. In these underserved communities, it is often women and children that are the recipients of non-profit healthcare. Impoverished communities are among the highest benefactors of ICT systems such as mobile applications that offer information and advice on how to treat symptoms without having to travel long distances. Such healthcare accessibility is an important component of sustainable development in such communities (Heeks 2017, Qureshi 2009, Qureshi & Xiong 2021).

Two communities in Omaha exhibit the social inequities that affect health disparities. Hispanic communities in South Omaha are close-knit and widely supportive of each other. The community is limited by healthcare expenses, of which are beyond the economic means of the majority in the area. This environment is a prime scenario for the insertion of ICTs. The African American communities in North Omaha, though similar in low economic development, is dissimilar in the community-supporting factors that South Omaha exhibited. The neighborhoods in this region were lacking evident forms of financial support; this was noted by directly observing the conditions of the infrastructure and conditions of the community in general (Qureshi 2020).

## 2. METHODOLOGY

This study follows the design science approach (Table 1). Hevner et al. (2004) assert that information systems (IS) exist as a discipline designated to improve effectiveness and efficiency of organizations. To do so, a theoretical framework of an existing problem must be determined (behavior science) and the process of effectively addressing that problem must be established (design science). Hevner et al. focus on the role of design science while paying tribute to the necessity of behavioral science, and lays out a framework to guide researchers and practitioners of IS on "how to conduct, evaluate, and present design science research (Hevner et al., 2004)."

| **Guideline** | **Hevner's Description** | **Our Project** |
|---|---|---|
| Guideline 1: Design as an Artifact | Design-science research must produce a viable artifact in the form of a construct, a model, a method, or an installation | IOS Application/ Website |
| Guideline 2: Problem Relevance | The objective of design-science research is to develop technology-based solutions to important and relevant business problems | The problem: The barrier between knowledge of community resources and members of the community who would benefit from these resources. The app/website addresses this problem. |
| Guideline 3: Design Evaluation | The utility, quality, and efficacy of a design artifact must be rigorously demonstrated via well-executed evaluation methods | Interviews will be conducted to address the utility, quality, and efficacy of the design artifact (app/website) |
| Guideline 4: Research Contributions | Effective design-science research must provide clear and verifiable contributions in the area of the design artifact, design foundations, and/or design methodologies | Our design is a "one stop shop" for resources because most people who need one resources would benefit from multiple resources. The website is condensed to address existing community health needs and available resources that offer low cost and no cost services (as money is a barrier for many). |
| Guideline 5: Research Rigor | Design-science research relies upon the application of rigorous methods | Data analysis from website visits, feedback survey, interviews |

| | in both the construction and evaluation of the design artifact. | |
|---|---|---|
| Guideline 6: Design as a Search Process | The search for an effective artifact requires utilizing available means to reach desired ends while satisfying laws in the problem environment. | Focusing on low cost/free services keeps the website relevant to the needs of the community. A 6$^{th}$ grade or lower reading level makes the website accessible to more people regardless of educational attainment or proficiency in English. Organization of information allows users to home in on specific types of resources rather than needing to look at specific providers and waste time on various website to see if the resource is relevant to their needs or not. |
| Guideline 7: Communication of Research | Design-science research must be presented effectively both to technology-oriented as well as management-oriented audiences. | Write up and presentation of our findings. |

**Table 1 Below is a table that describes shows how Hevner et al.'s guidelines apply to our project.**

The relationship between behavioral science and design science creates what the authors call a "build-and-evaluate loop," in which a problem is determined, a solution is designed, the solution is evaluated, and the solution is tweaked until its solution is effective. This will be addressed through interviews and analysis of website data.

The action research approach was taken in this study because of its demonstrated ability to research the "conditions and effects of various forms of social action," and with that research, lead to social action (Lewin, 1946). The process followed what we term "IT therapy" as it involves a set of steps:

1.  **Diagnosis of problems and needs assessment:** This involves understanding the development context in which the clinic operates. When people cannot afford to pay for the healthcare they need, they may not be able to follow through with the treatments that are recommended to them, although assistance can be granted upon appeal. That is why it is important to identify the socio-economic factors affecting the people who seek care at this clinic.
2.  **Identification and trial of alternative solutions:** This involves finding context-appropriate solutions to problems that were diagnosed. The criteria for selection of alternatives involves solutions that are free or cost effective, take little time to diagnose, and are easy to use.
3.  **Development and implementation of IT solution:** The most appropriate IT solution is one that fits the criteria identified above and is quickly implementable. A prototyping method is used here to ensure that the users can offer feedback on improvements to the mobile app while using it to address the socio-economic needs of their patients.
4.  **Adoption and use of IT solution:** Observations are made of the use and changes in usage patterns of the mobile app. They are recorded and analyzed for appropriateness to the development context.
5.  **Sustainable Development Outcomes and Community Impact:** Outcomes to sustainable development are identified in terms of human development outcomes: what learning, and skill development has taken place because of the interventions?

The researchers ventured into the regions of North and South Omaha to investigate the development factors that were exhibited in each community. These important observations were necessary to evaluate and initiate a student-led implementation of an ICT that would contribute to the community. The implementation would benefit a microenterprise or non-profit on the micro-level and help facilitate the sustainable development cycle at the root. The following sections report on the interventions and offer observations of the interventions.

# 3. INTERVENTION

Addressing the needs of marginalized communities requires interventions that are culturally sensitive while understanding the socio-economic causes of the patient's illness. The free Student Health Alliance Reaching Indigent Needy Groups (SHARING) clinic opened on September 9, 1997 in order to provide primary health care to underprivileged populations in South Omaha (University of Nebraska Medical Center). The SHARING clinic now operates at the junction between the North and South Omaha communities. The clinic lacked financial resources; therefore, a free solution was implemented that fulfilled their need to offer a growing body of services. A mobile application with information on community resources was developed so that the clinicians could find relevant socio-economic information for their patients. To offer clinicians with the ability to address the socio-economic needs of their patients, a series of interventions comprising of technology and training solutions were carried out at the free student clinic. The Free Student Led Clinic was started in 1997 to provide primary health care to multi-ethnic underserved populations in North and South Omaha. Patients who qualify for care at this clinic do not have health insurance and are living in poverty. The clinic is located in the middle of North and South Omaha serving the African American and Hispanic populations.

## 3.1. Diagnosis of problems and needs assessments.

While clinicians can address the symptoms that patients present with, longer term health of their patients need to be addressed based on the social and economic causes of their illness. The purpose of the mobile app is to: 1) allow clinicians to look up appropriate community resources based on patient needs, 2) allow patients to leave the clinic with a plan for further support, and 3) allow clinics to help our patients navigate bio-psycho-social barriers more meaningfully. Ten clinicians working in the clinics were interviewed about the requirements for the application to determine their information needs and to get some insight into the organization's culture. A list was created of the information the clinicians would like in the application: 1) List of community resources 2) Nutrition guidelines documentation 3) Food, Dental, and Transportation related information 4) Scheduling system integration 5) Materials which can help to deal with a difficult conversation and sensitive cultural topics. Their top priority was an application to provide their patients with community resources. Some challenges mentioned during the interviews were being pressed for time, money, and resource constraints.

## 3.2. Identification and trial of alternative solutions

After gathering the requirements from the clinicians, the knowledgebase was created. The knowledge base provides a Cloud Solution for Community Resources. The Google Cloud Platform was used, which is a suite of cloud computing services that runs on the same infrastructure that Google uses internally for its end-user products, such as Google Search and YouTube. Google's cloud platform provides a reliable and highly scalable infrastructure for developers to build, test, and deploy apps. It covers application, storage, and computing services for backend, mobile, and web solutions. More than four million apps trust and use the platform (Rohit Dogra, 2014).

## 3.3. Development and implementation of IT solution

To implement the free service, researchers put together a Google Sites application that fulfilled the community resources requirements. This is available as a website (Figure 1) for clinicians who are at their desks as well as a mobile application (Figure 2) that can be accessed through their phones. Researchers and clinicians walked through a prototype of the app to determine if the app met their needs.

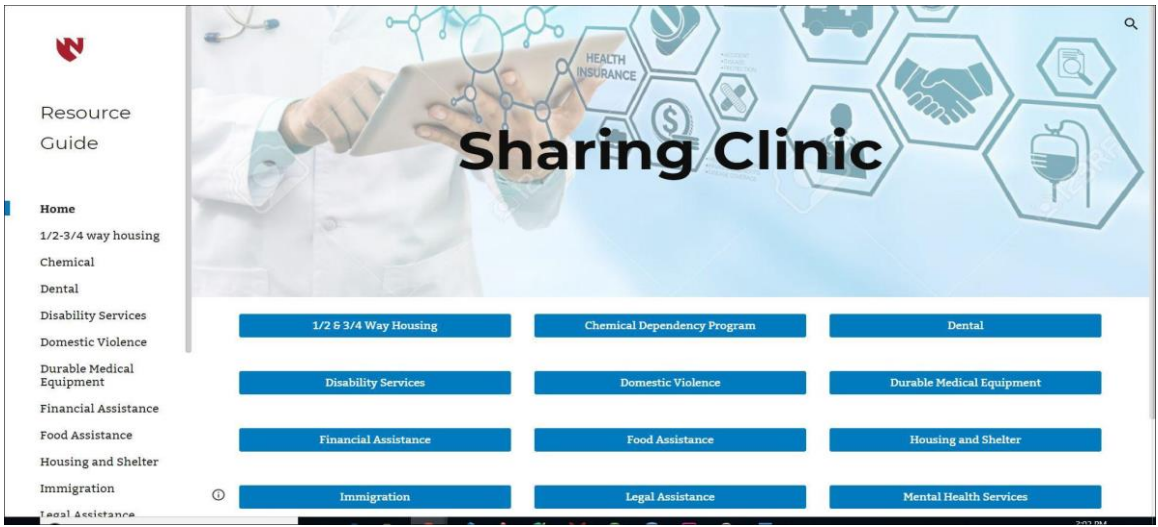

**Figure 1 Screenshot of Website home screen**

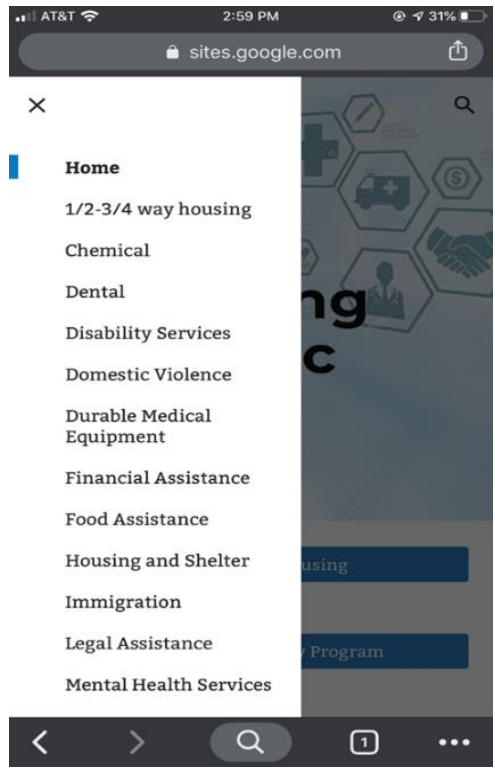

**Figure 2 Screenshots of Mobile App main screen**

## 4. RESULTS

### 4.1. Adoption and use of IT solution

To ensure the sustainability of the mHealth intervention, the researchers observed how the app was being used by clinicians within the cultural context of the clinic (Krauss 2013). The clinic provides the opportunity to have all their patients' health care needs met, such as, access to primary care, dental care, and vision care, and they often must rely on low-cost community resources to fulfil these needs. The observations below illustrate this need:

When resources recommended how is it shared? Verbally? Written? How do patients remember? Is there a follow up with patients? i.e., "did you visit _______?"

> Concern: if requirements not available on app, may lead to medical clinicians/staff recommending irrelevant information/resources
>
> If patients given "homework" and left overwhelmed- (resources relevant to their specific needs are difficult to locate even after referral) they may give up or lose hope- may be difficult to research without speaking with patients

### Transcript 1 Initial Description of Requirements

The application allows clinicians to provide immediate referral during consultation, which allows them to provide better and more informed care for their patients. Furthermore, clinicians who use the app can reduce the amount of time spent searching for community resources, and resource eligibility requirements. The following observation relates to how this app is used by clinicians during consultation:

> App recommendations after speaking with Dr. M (Social Services) Accessibility- keep it as disability services.
>
> Clinician B: "Do you want to flip to it?"
>
> M: "How do I do that? I don't have an iPad." Make app user friendly for users of all levels of technological use knowledge. Don't need physical health in app- that is what they do at Clinic

### Transcript 2: Use during Consultation

The use of the mobile app offered a unique learning opportunity for the student clinicians in several ways. First, the app introduced the student clinicians to various community resources, which they will be able to refer to in their future practice. Second, it allowed the students to reflect on the implications of health for underserved communities and socio-economically vulnerable populations. The following transcripts illustrates how the app was used by clinicians to address the socioeconomic needs of patients:

> Dr. M: when looking through resources "Mental health is good." In Douglas County- Keep Chemical dependency- move it up to mental health. (list related services within close proximity to one another on app) i.e., mental health, and chemical dependency

### Transcript 3: Addressing socio-economic needs of patients

Offering resources to their patients may improve the clinician-patient relationship in that the patient will be assured that their clinician is listening to their needs and attempting to remedy them; even in areas that are outside of the clinician's immediate realm of practice. The following observations illustrate how the app is used to address these unique needs:

> Add community support to disability services
>
> Neb. Med Financial Assistance
>
> [Patient comes in with "strange" characteristics] as noted by front desk workers. Dr. M notes they are symptoms of a mental illness. States that it is good for medical professionals to spend time working in an in-patient psych ward to be able to identify these characteristics.] How could they have the ability to identify these characteristics improve patient health? Patients might be confused or unsure of what they need.

### Transcript 4: Sensitivity to patient needs

These results suggest that to remain healthy, people from low socio-economic backgrounds will need access to the socio-economic resources available through the mobile application to stay healthy.

# 5. ANALYSIS

## 1.1. Sustainable Development Outcomes and Community Impact

To achieve sustainability of the technology and training intervention, clinicians need to be able to use the app to support the socioeconomic needs of their patients. Following the initial implementation of the app, the researchers visited the student-led free clinic on four separate occasions to observe how the app was being used by the clinicians. Specifically, the researchers sought to analyze how the app was being used to address patient needs, how the app affected patient access to resources, how the app could be improved to ensure the relevancy of its resource list, and ensure that the app was considered user friendly among clinicians. This analysis is offered in this section.

**Use of App to address patient needs.** One clinician, a medical student, informed the researchers that he had a particular interest in understanding and addressing the socioeconomic circumstances that may influence one's health outcomes, and he expressed enthusiasm in the app's resources as the services listed confront the most pressing of needs met by his patients, such as financial assistance, mental health services, and dental care. Other clinicians noted that the app is relatively user friendly, even among clinicians who expressed lower confidence in their abilities to operate mobile technologies. The following transcript illustrates how the app is used to address patients' socio-economic needs:

> When looking at app updates. Dr. M. "Wow that is so cool. This is nice."
>
> Is there a way to print or copy information to give it to patients? - Asked by Clinician A, a student. Clinician A also expressed pleasure that the phone numbers for the services was included. "Phone numbers. Nice."
>
> On the app- "Pretty smart. Pretty good looking." Clinician A
>
> When told that the app can be opened on multiple devices (i.e., cell phone, iPad, desktop) "nice. That's convenient. Cool." -Clinician A
>
> Clinicians using the app noted that it was convenient and easy to use.
>
> Dr. M expressed pleasure in that you can print off information from the app. "Wow. Look at that."
>
> Clinician D, to himself while looking through the app. "Oh man." -tone appears excited. "Right on. I like this. "[The app is] very straightforward," Clinician D, to Researcher P and Researcher M. "I like how easy this looks right at the beginning. Clinician D states that he would rate it a 5 when prompted to rate the app.

<div align="center">Transcript 5: Addressing patient needs using the app</div>

**Access to resources.** Providing access to socioeconomic resources in the community are important for the patients. Yet, if clinicians can access these, the chances of the patients being able to avail these are higher. The mobile app offers this information to clinicians. Their use is described in the following transcript:

> Dr. H. appears to have learned more about housing services through the app, "Oh, there's a lot of places for housing [in Omaha]."
>
> While looking through the app. Dr. H. "That's great. Wow. That's a lot of stuff."
>
> "Do you have that central building [with all the] services? A one stop shop?"- Dr. H
>
> After speaking with Dr. M about Dr. H comment, she informed us that such a service does not exist. Dr. M suggested that Dr. H may be thinking of a multi-service organization. Perhaps "Multi Service Organizations" could/should be a section on the app.
>
> Dr. M. Expressed interest in using the app for the Respect Clinic as well.
>
> Clinician B, a student, is using the app to pull information for a patient.

Clinician B, while looking through the app. "Oh. That's really cool."

Clinician D feels that the app could save the users time and that it is relevant to the services their patients are looking for

Clinician D mentioned that he is learning more about underserved communities in the medical field and that this app covers a lot of the resources underserved communities will need.

### Transcript 6: Accessing resources for patient needs

**Improvements to ensure the relevancy.** The lead faculty member for social work at the student-led free clinic served as a key informant for the app's content and regularly offered feedback regarding the resource list and its relevance to the population seen at the clinic. While the initial version of the app included resources that were deemed necessary, many of the resources were deemed obsolete because of their location, their health insurance requirements, or their wealth requirements. It was suggested that all resources outside of the immediate vicinity of the clinic be removed as most of their patients are locals and will not travel, or do not have the necessary means, to travel for resources. Additionally, any resource that required proof of insurance, or did not offer low-income services was removed from the app and replaced with more appropriate resources because the patients of the clinic are low income and many do not have health insurance coverage. These are indicated in the following transcript:

"Is that [Moby] the only transportation?"- Dr. H. Perhaps add more transportation if there is any

Description for ¾ way house was confusing for Dr. H. Perhaps simply keep it under housing services After speaking with Dr. M about Dr. H comment, she informed us that such a service does not exist. Dr. M suggested that Dr. H may be thinking of a multi-service organization. Perhaps "Multi Service Organizations" could/should be a section on the app.

Dr. M suggests under-lining titles and using single space for ease of use. Double space between services. "We want it to be super easy on the eyes" Remove any services that are not in Omaha Our people are local." "We need to add vision on here."-

Dr. M. "The layout throws me off." – Dr. M. Then repeats her suggestion of using single space, underlining titles, and only using double space between listed services. She stated that the clinic does not collect socio-economic data. However, they do have guidelines that clients must meet before being allowed services. Requirements included on the last page of this document

Clinician D, a student, stated that the URL is long and suggested shortening it if possible. (If not possible perhaps providing a link for all clinicians would be easier)

### Transcript 7 Improvements to ensure the relevancy

Various clinicians also offered feedback regarding ease of use and impressions of the app's implementation process. They noted that the layout of the app was initially confusing. After further discussion and clarification, a redesign of the app layout was implemented. For example, titles for individual resources were underlined and double spacing was used to signify when a new resource was listed. Clinicians agreed that the update improved the overall readability of the app.

COVID-19 is a global health crisis increasing human suffering, destabilizing the global economy and unsettling the lives of people around the world. As COVID-19 struck shortly after the app was implemented, the free clinics were shut down. This meant that the community resources made available through the app were updated to include COVID-19 related resources. Since COVID-19 struck, the mobile application was made available to the public and is being used to support the individuals suffering from the adverse effects of the shutdowns. Recent data suggests that among the effects of the COVID-19 shutdown, there have been increases in domestic violence, homelessness, and food insecurity. There has also been a sudden rise in the cost of prescription drugs. We expect that with the resources made available through this app, such hardship can at least, to a certain extent, be alleviated for those who are able to access the resources (Qureshi, 2020).

ICTs can address sustainable development, especially in urban settings by ensuring that underserved groups have access to information on resources needed to survive during the pandemic. Providing access to socioeconomic resources in the community are important for the patients. Understanding and addressing the socioeconomic circumstances that may influence a patient's health outcomes is a step toward meeting patients' needs, such as, financial assistance, mental health services, and dental care. Ensuring relevancy by providing resources based on patients' location, health insurance requirements, or their wealth requirements is necessary to increase usefulness of the app.

## 6. SUMMARY AND CONCLUSIONS

Ensuring healthy lives and promoting well-being is necessary for sustainable development. This paper reports on a study of how technology and training interventions can enable clinicians in a student led free clinic to offer socio-economic based care to their marginalized patients. The interventions allowed clinicians to search for appropriate community resources based on patient needs and allowed patients to leave the clinic with a plan for further support, which helped the patients navigate the socio-economic barriers. Clinicians were able to use the app to better manage the care of their marginalized patients by providing them with information on community resources. Through sets of app revisions and observations of usage, this paper arrives at insights into how such applications can support multi-ethnic underserved communities.

Our research was influenced by the pandemic in diverse ways. Conducting fieldwork in the form of ethnography became impracticable because of the hospital shutdown of all human subjects research and individuals that were not a part of the healthcare team were unable to access the facilities. Also, the SHARING clinic was shut down completely, no longer accepting patients, which compelled us to resort to alternative ways to provide patients with access to the information they needed. Resilience became crucial in the conduct of our research. Resilience is "the ability of systems to cope with external shocks and trends (Heeks and Ospina 2019)." As a result of COVID-19, the SHARING clinics were shut down and a mobile application was developed to provide information on community resources to patients of this clinic who are unable to access health resources. Although the clinic was shut down, we were able to change the scope of our research by making the app available for public use instead of just a clinicians resource so patients are able to access our app to find the resources they need to lead healthy lives.

## REFERENCES

Adler, N. E., Boyce, T., Chesney, M. A., Cohen, S., Folkman, S., Kahn, R. L., & Syme, S. L. (1994). Socioeconomic status and health: The challenge of the gradient. American psychologist, 49(1), 15.

Caucutt, N. Guner, C. Rauh (April, 2019). Incarceration, Unemployment, and the Black-White Marriage Gap in the US. Vox CEPR Policy Portal. Retrieved from https://voxeu.org/article/incarceration-unemployment-and-black-white-marriage-gap-us

Clarke, M. A., Lyden, E. R., Ma, J., King, K. M., Siahpush, M. M., Michaud, T., ... & Ramos, A. K. (2020). Sociodemographic differences and factors affecting patient portal utilization. Journal of racial and ethnic health disparities, 1-13.

Office of Assistant Secretary for Planning and Evaluation. 2020 Poverty Guidelines. (2020). Retrieved 20 May 2021, from https://aspe.hhs.gov/2020-poverty-guidelines

Hevner, A. R., March, S. T., Park, J., Ram, S., & Ram, S. (2004). Research Essay Design Science in Information. MIS Quarterly, 28(1), 75–105. https://www.jstor.org/stable/25148625

University of Nebraska Medical Center. Student Health Alliance Reaching Indigent Needy Groups (SHARING). Retrieved from https://www.unmc.edu/sharing/clinics/sharing-clinic.html

Lewin, K. (1946). Action Research and Minority Problems. Journal of Social Issues, 2(4), 34–46. https://doi.org/10.1111/j.1540-4560.1946.tb02295.x

Jones, M. (December, 2018). HealthCare: How Technology Impacts the Healthcare Industry. Healthcare in America. Retrieved from https://healthcareinamerica.us/healthcare-how-technology-impacts-the-healthcare-industry-b2ba6271c4b4

Qureshi, S. (2009) Social and economic perspectives on the role of information and communication technology for development, 15:1, 1-3, DOI: 10.1002/itdj.20117

Qureshi, S. (2020). Bridging Socio-Economic Inequities in Healthcare Access: A Mobile Health Based Approach. IFIP 9.4 Blog https://ifip94.wordpress.com/2020/05/07/bridging-socio-economic-inequities-in-healthcare-access-a-mobile-health-based-approach/

Qureshi, S., & Xiong, J. (2021). Equitable healthcare provision: uncovering the impact of the mobility effect on human development. Information Systems Management, 38(1), 2-20.

Ramos, A. K., Carvajal-Suarez, M., Siahpush, M., Robbins, R., Michaud, T. L., Clarke, M. A., & King, K. M. (2020). Predictors of life satisfaction among Hispanic/Latino immigrants in non-metropolitan communities in the Midwest. Rural Society, 29(2), 75-88.

Roztocki, N. & H. Roland Weistroffer (2016) Conceptualizing and Researching the Adoption of ICT and the Impact on Socioeconomic Development, Information Technology for Development, 22:4, 541-549, DOI: 10.1080/02681102.2016.1196097

Siahpush, M., Robbins, R. E., Ramos, A. K., Michaud, T. L., Clarke, M. A., & King, K. M. (2019). Does Difference in Physical Activity Between Blacks and Whites Vary by Sex, Income, Education, and Region of Residence? Results from 2008 to 2017 National Health Interview Surveys. Journal of racial and ethnic health disparities, 6(5), 883-891.